\documentclass[fleqn,12pt,twoside]{article}
\usepackage[headings]{espcrc1}

\readRCS
$Id: espcrc1.tex,v 1.2 2004/02/24 11:22:11 spepping Exp $
\usepackage{graphicx}
\usepackage[figuresright]{rotating}

\newcommand{\AmS}{{\protect\the\textfont2
  A\kern-.1667em\lower.5ex\hbox{M}\kern-.125emS}}

\title{News about TeV-scale black holes}

\author{S. Hossenfelder\address{Department of Physics, University of Arizona\\
1118 E 4th Street, Tucson, AZ 85721, USA}%
        \thanks{sabine@physics.arizona.edu}}
       
\runtitle{News about TeV-scale Black Holes}
\runauthor{S. Hossenfelder}

\begin{document}

% typeset front matter
\maketitle

\begin{abstract}
Collider produced black holes are the most exciting prediction from models with large extra dimensions. 
These black holes exist in an extreme region, in which gravity meets quantum field theory, particle physics, and thermodynamics. 
An investigation of the formation and decay processes can therefore provide us with important insights about the 
underlying theory and open a window to the understanding of Physics at the Planck scale.  
The production and the evaporation of TeV-scale black holes yields distinct signatures that have been examined closely 
during the last years, with analytical approaches as well as by use of numerical simulations. 
I present new results for the 
LHC, which take into account that, instead of a final decay, a black hole remnant can be left. 
\end{abstract}

\section{Black Holes in High Energy Collisions}

High energetic particle collisions will eventually lead to strong
gravitational interactions and result in the formation of a black hole's horizon.
In the presence of large additional
compactified dimensions \cite{Arkani-Hamed:1998rs}, it could be 
possible that the threshold for black
hole production lies within the accessible range for future experiments.
In the context of models with such extra dimensions, the black hole 
production is predicted to drastically change high energy physics at the LHC.
These effective models are  
string-inspired \cite{Antoniadis:1990ew,Antoniadis:1996hk,Dienes:1998vg} 
extensions to the Standard Model in the overlap region between
'top-down' and 'bottom-up' approaches.  

The possible production of
TeV-scale black holes at the {\sc LHC} is
surely one of the most exciting predictions of physics beyond
the Standard Model and has
received a great amount of interest during the last years.
For reviews on the subject the interested reader is referred to \cite{Kanti:2004nr}.

Due to their Hawking-radiation, these black holes have an high
temperature of some $100$~GeV and decay very fast into $\sim 10-25$ thermally 
distributed particles of the Standard Model (before fragmentation), which yields a 
signature unlike all other new predicted effects. 
The black hole's evaporation process connects quantum gravity with
quantum field theory and particle physics, and is a promising way towards the
understanding of Planck scale physics.

Black holes are a fascinating field of research which features an interplay
between General Relativity, thermodynamics, quantum field theory, and
recently also particle physics. The investigation of black holes objects  would
allow us to test Planck scale effects and the onset of quantum gravity. 
The understanding of the black holes properties thus
is a key knowledge to the phenomenology of physics
beyond the Standard Model. 
 
Recently, the production of black holes has been incorporated into 
detailed numerical simulations of black hole events at the {\sc LHC} and
their detection \cite{Atlas}. These important investigations
allow us to reconstruct initial parameters of the model from observed data.

So far these simulations have assumed that the black hole decays 
in its final phase completely into some few particles of the Standard Model. 
However, from the theoretical point of view, there are strong indications that the black
hole does not evaporate completely, but leaves a stable black hole remnant. 
In a recent work, we included this possibility into the
numerical simulation and examined the consequences for the observables 
of the black hole event \cite{Koch:2005ks}.

\section{Signatures of Black Hole Relics}

We have parameterized the modifications to the black hole evaporation arising from the
presence of a remnant mass and included these modifications in a numerical simulation for
black hole events at the LHC, for details see \cite{Koch:2005ks}. The relevant parameter
is the mass of the remnant, $M_{\rm R}$ that we assume to be close by the new fundamental
scale $M_{\rm f}$.

In the regime of interest here, when the mass of the black hole,
$M$, is of order $M_{\rm f}$, the emission
rate for a single particle microstate has to take into account the backreaction on
the black hole and is given by 
\begin{equation} \label{nsingle}
n(\omega) =  \frac{\exp[S(M-\omega)]}{\exp[S(M)]}\quad.
\end{equation}
where $S$ is the black hole's entropy.

For the spectral energy density we then use this particle 
spectrum and integrate over the momentum space. From this one obtains the
evaporation rate with the Stefan-Boltzmann law to
\begin{eqnarray} \label{mdoteq}
\frac{{\mathrm d}M}{{\mathrm d}t} = \frac{\Omega_{(3)}^2}{(2\pi)^{3}} R_H^{2} 
\int_0^{M-M_{\rm r}} \hspace*{-2mm}  
\frac{\omega^3 ~{\mathrm d}\omega }{{\rm exp}[S(M-\omega)-S(M)] + s}   \quad . 
\label{dmdtrel}
\end{eqnarray}
where $R_H$ is the Schwarzschild-radius of the black hole and typically of order $1/M_{\rm f}$, and
$s$ labels the spin-statistic of the emitted radiation.
The appearance of the remnant as a smallest possible mass is captured in the
upper integration bound. 

%..........................................................................
\begin{figure} 

\vspace*{-1.5cm}
\hspace*{0cm} 
\includegraphics[width=14cm]{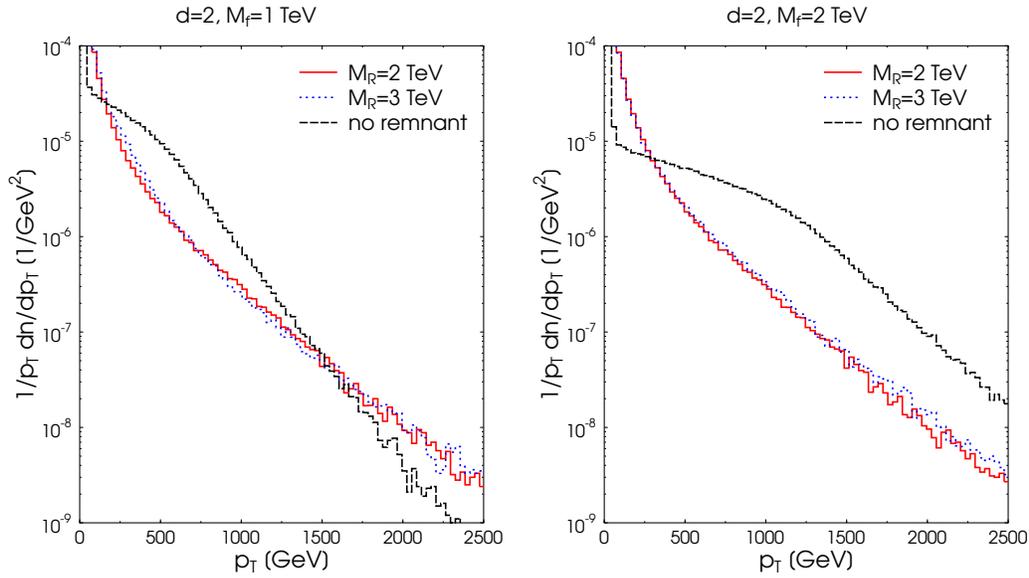}

\vspace*{-0.3cm}
\caption{Transverse momentum distribution of  initially emitted
particles (before the fragmentation of the emitted partons)
with final (two body) decay in contrast to the formation of a black
hole remnant. \label{fig8}}
\end{figure}
%..........................................................................

%..........................................................................
\begin{figure}

\vspace*{-3.0cm} 
\includegraphics[width=14cm]{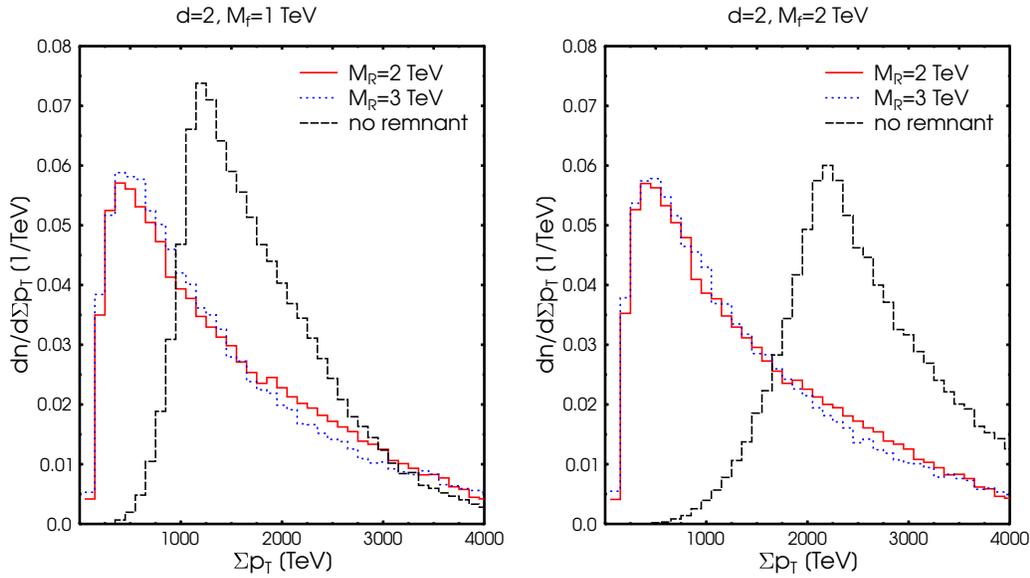}

\vspace*{-0.3cm}
\caption{The total sum of the transverse momenta of the decay
products. \label{fig11}}

\end{figure}
%..........................................................................

For the
present examination we have initialized a sample of 50,000 events
of  black hole
remnants in a proton-proton collision at $\sqrt{s}=14$~TeV.
 All plots are for $d=2$ since a higher number 
of extra dimensions leads to variations by less than 5\%. 
The produced black hole remnants are strongly peaked around central rapidities,
making them potentially accessible to the CMS and ATLAS experiments.

Figure \ref{fig8} shows the transverse momentum, $p_T$, of the decay
products as it results from the modified evaporation rate Eq. (\ref{dmdtrel}) before fragmentation. One
clearly sees the additional contribution from the final decay which
causes a bump in the spectrum that is absent in the case of a
remnant formation. An examination of the sampled data shows that after fragmentation, 
this bump is slightly washed out but still present. However, from the rapidity distribution and
the fact that the black hole event is spherical, a part of the high
$p_T$-particles will be at large $y$ and thus be not available in
the detector. We therefore want to mention that one has to include
the experimental acceptance in detail if one wants to compare to
experimental observables.

Figure \ref{fig11}  shows the sum over the transverse momenta of the
black hole's decay products. To interpret this observable one might
think of the black hole event as a multijet with total $\Sigma p_T$.
As is evident, the formation of a remnant lowers the total $p_T$ by
about $M_{\rm R}$. This also means that the signatures of the black
hole as previously analyzed are dominated by the assumed final and
not by the Hawking phase.

 An examination of the total
multiplicities of the event shows that the formation a black hole remnant
increases the multiplicity due to the additional low energetic
particles that are emitted in the late stages instead of a final
decay with $2-5$ particles. 

It is interesting to note that the dependence on $M_{\rm f}$ is
dominated by those on $M_{\rm R}$, making the remnant mass the
primary observable.

\section{Conclusions}

We examined the formation of black hole remnants  in  proton proton collision in contrast
to a final decay and found
the total multiplicity of the event to be lowered by $\sim$~100 and the missing 
transverse momentum to be increased.  
We have shown that the formation of black hole remnants in high energy collisions
would yield signatures that differ significantly from the total decay of black holes.

\section*{Acknowledgments}
 This work was supported by the  Frankfurt Institute of Advanced Studies
({\sc FIAS}). SH thanks Marcus Bleicher, Ben Koch and Horst St\"ocker for the
fruitful collaboration.

\end{document}